\documentclass[%
 aip,
amsmath,amssymb,
preprint,%
]{revtex4-1}
\usepackage{color}%
\usepackage{graphicx}%
\usepackage{dcolumn}%
\usepackage{bm}%
\usepackage{natbib}

\draft 

\begin{document}


\title{Transition from ion-coupled to electron-only reconnection: {B}asic physics and implications for plasma turbulence} 




\author{P. Sharma Pyakurel}
 \affiliation{Department of Physics and Astronomy, University of Delaware, Newark, Delaware 19716, USA}

\author{M. A. Shay}%
\affiliation{Department of Physics and Astronomy, University of Delaware, Newark, Delaware 19716, USA
}%

\author{T. D. Phan}
\affiliation{%
Space Sciences Laboratory, University of California, Berkeley, CA  94720, USA
}%

\author{W. H. Matthaeus}%
\affiliation{Department of Physics and Astronomy, University of Delaware, Newark, Delaware 19716, USA
}%

\author{J. F. Drake}
\affiliation{%
Department of Physics and the Institute for Physical Science and Technology, University of Maryland, College Park, MD 20742, USA
}%

\author{J. M. TenBarge}
\affiliation{%
Department of Astrophysical Sciences, Princeton University, Princeton, NJ 08544, USA
}%

\author{C. C. Haggerty}
\affiliation{%
Department of Astronomy and Astrophysics, University of Chicago, Chicago, IL, 60673, USA
}%

\author{K. Klein}
\affiliation{%
Lunar and Planetary Laboratory, University of Arizona, Tucson, AZ 85719, USA
}%

\author{P. A. Cassak}
\affiliation{%
Department of Physics and Astronomy, West Virginia University, Morgantown, WV 26506, USA
}%

\author{T. N. Parashar}%
\affiliation{Department of Physics and Astronomy, University of Delaware, Newark, Delaware 19716, USA
}%

\author{M. Swisdak}
\affiliation{%
Department of Physics and the Institute for Physical Science and Technology, University of Maryland, College Park, MD 20742, USA
}%

\author{A. Chasapis}%
\affiliation{Department of Physics and Astronomy, University of Delaware, Newark, Delaware 19716, USA
}%

\date{\today}

\begin{abstract}
Using kinetic particle-in-cell (PIC) simulations, we simulate reconnection conditions appropriate for the magnetosheath and solar wind, i.e., plasma beta (ratio of gas pressure to magnetic pressure) greater than 1 and low magnetic shear (strong guide field). Changing the simulation domain size, we find that the ion response varies greatly. For reconnecting regions with scales comparable to the ion Larmor radius, the ions do not respond to the reconnection dynamics leading to ``electron-only'' reconnection with very large quasi-steady reconnection rates. The transition to more traditional ``ion-coupled'' reconnection is gradual as the reconnection domain size increases, with the ions becoming frozen-in in the exhaust when the magnetic island width in the normal direction reaches many ion inertial lengths. During this transition, the quasi-steady reconnection rate decreases until the ions are fully coupled, ultimately reaching an asymptotic value. The scaling of the ion outflow velocity with exhaust width during this electron-only to ion-coupled transition is found to be consistent with a theoretical model of a newly reconnected field line.  In order to have a fully frozen-in ion exhaust with ion flows comparable to the reconnection Alfv\'en speed, an exhaust width of at least several ion inertial lengths is needed. In turbulent systems with reconnection occurring between magnetic bubbles associated with fluctuations, using geometric arguments we estimate that fully ion-coupled reconnection requires magnetic bubble length scales of at least several tens of ion inertial lengths.
\end{abstract}

\maketitle 

\section{Introduction}
Magnetic reconnection is a magnetic energy release process that plays a fundamentally important role in laboratory, space, and astrophysical plasmas~\cite{Yamada10}. The role that magnetic reconnection plays in damping turbulent fluctuations in plasma has significant implications for our understanding of diverse systems such as the solar corona, the solar wind, the Earth's magnetosheath, and astrophysical accretion disks. While magnetic reconnection has been observed in the turbulent magnetosheath of the Earth~\cite{Retino07,Yordanova16,eriksson16,Voros17,Phan18}, our understanding of the role it plays in damping turbulent magnetic energy and heating the plasma is incomplete. Two-dimensional magnetohydrodynamics (MHD) simulations and Hall MHD simulations of turbulence have been used to study the statistics of reconnection, finding a large spread of reconnection rates at x-lines occurring as part of the turbulence{~\cite{Servidio10, Donato2012}}. The x-lines showing robust reconnection had reconnection rates consistent with quasi-steady theories of reconnection~\cite{Cassak07d}. Recently, these x-line finding techniques were applied to fully kinetic simulations of turbulence~\cite{Haggerty17}, where a similar spread of reconnection rates was found. The effect of reconnection on the cascade of energy and even as a driver of the cascade has recently been the focus of significant scrutiny~\cite{Cerri17,Dong2018,Mallet17,Boldyrev17,Franci17,Papini18}. A framework for estimating the heating due to reconnection in turbulence has been established~\cite{Shay18}, which draws on recent studies of heating during isolated laminar reconnection~\cite{Phan13,Phan14,Shay14,Haggerty15}. 

\begin{figure*}
\centering
\includegraphics[width=0.88\columnwidth]{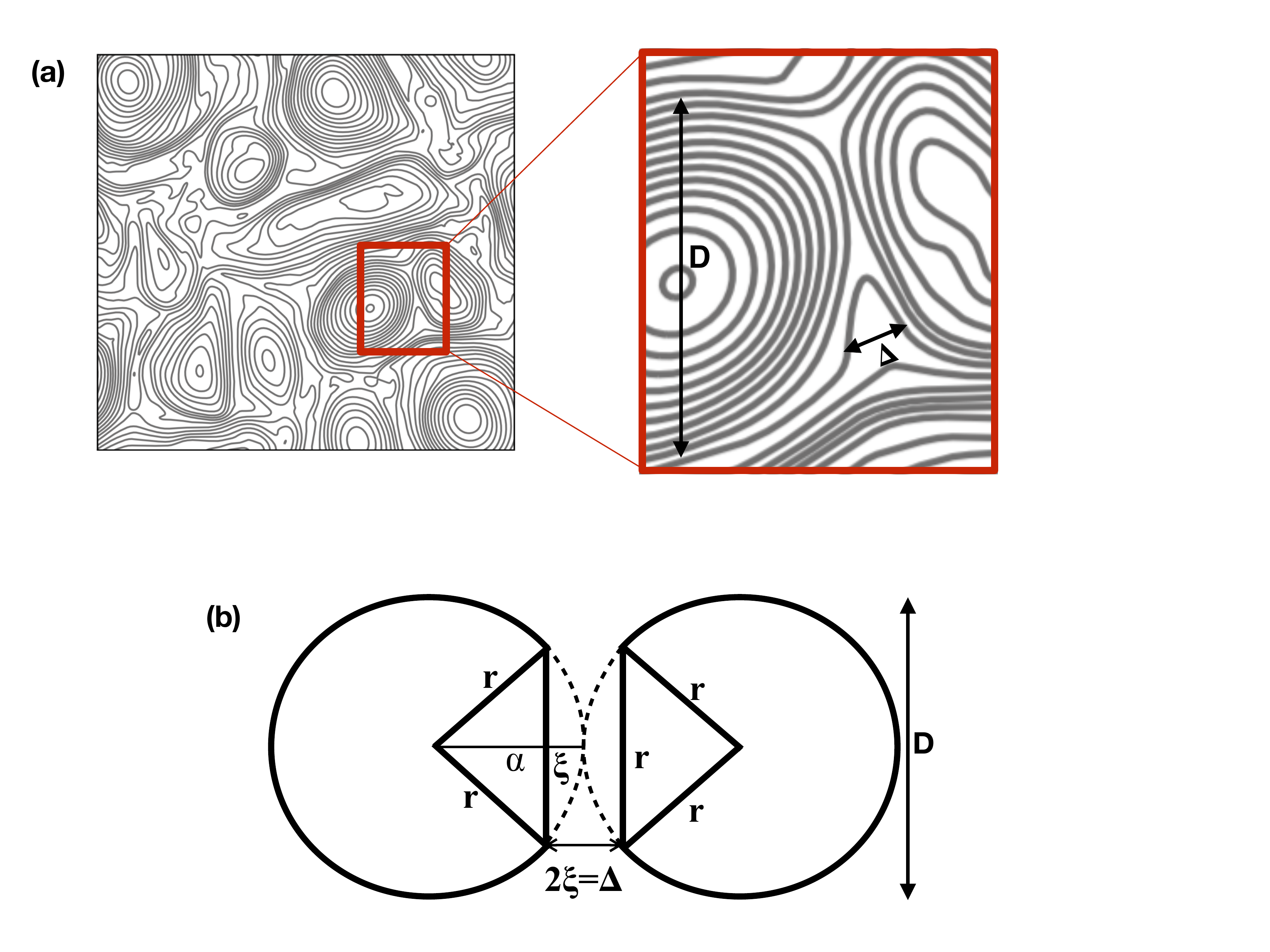}
\caption{
(a) Schematic of magnetic field lines adapted from ~Ref.~\onlinecite{Phan18} showing an enlargement in the vicinity of a magnetic reconnection region. Shown are the approximate exhaust width $\mathrm{\Delta}$ from the reconnection of a magnetic bubble roughly of size $\mathrm{D}$. (b) Geometrical interpretation: two flux bubbles interact with radius $\mathrm{r}$ with a separation distance $\mathrm{\Delta}$. The figure is an illustration of bubble size threshold for the ions to respond to the reconnected field lines in magnetic reconnection. 
\label{fig:turb-schematic}}
\end{figure*}

In a low collisionality plasma, the cascade of turbulent energy from large energy containing scales to small scales raises the question as to the existence and properties of the magnetic reconnection at the smallest scales where turbulent energy is damped. At such small scales, it seems likely that reconnection may occur in a small enough region where the ions do not respond, i.e., ``electron-only reconnection'' occurs. In fact, recent observations of magnetic reconnection in the turbulent magnetosheath have observed magnetic reconnection occurring with no ion response~\cite{Phan18}.
 
Various aspects of electron-only reconnection have been studied previously
with both fluid and kinetic particle-in-cell (PIC) simulations~(e.g. Refs.~\onlinecite{Shay98a,Chacon07,Jain12}, and references therein). Simulation scaling studies~\cite{Biskamp95,Shay98a} found that the rate of quasi-steady reconnection is independent of the
electron mass. The decoupling of electron and ion velocities, e.g., Hall physics~\cite{Sonnerup79,Terasawa83}, was found to be a key factor in  this
independence. Studies of the transition from this Hall reconnection to more
typical ``ion-coupled reconnection'' have also been performed, showing that the timescale to reconnect flux transitions from a Hall timescale to one mediated by the MHD Alfv\'en time~\cite{Mandt94,Biskamp95}; note that we use the term ``ion-coupled'' to describe reconnection in which the ion outflow exhausts become frozen-in to the magnetic field. The transition between the Hall and MHD regimes occurred at simulation domain sizes of around 10 ion inertial lengths. Turbulence simulations driven at scales small enough so the ions are not coupled at the energy containing scale have found that in the vicinity of reconnection sites, electrons are preferentially heated in the direction parallel to the magnetic field~\cite{Haynes14}.

An important question concerning electron-only reconnection regards the limiting length scales and timescales for its existence. Magnetic reconnection in a turbulent system occurs between magnetic ``bubbles'' associated with the fluctuations in the magnetic field. A schematic of magnetic field lines in turbulence generated from a 2D turbulence simulation~\cite{Phan18} is shown in Figure~\ref{fig:turb-schematic}a. Two reconnecting magnetic bubbles (flux tubes in a 2D geometry) in a turbulent system are highlighted, and the approximate scale size $\mathrm{D}$ of a bubble is shown. The scale $\mathrm{D}$ is approximately the largest length scale associated with the reconnection and plays an important role in determining the degree of ion-coupling to the reconnection. This length scale is roughly equivalent to the simulation domain size of conventional simulations of laminar reconnection. Hence, simulating different domain sizes in laminar reconnection simulations can help shed light on the degree of ion coupling to reconnection in turbulence. 

Ultimately, at MHD lengths or
timescales, the reconnection must eventually couple to the ions. However,
previous simulations of this transition between ion-coupled and electron-only reconnection focused exclusively on the reconnection rate~\cite{Mandt94}. In addition, this study focused on low ion plasma $\beta$ and anti-parallel reconnection, whereas reconnection in the solar wind or Earth's magnetosheath is often characterized by strong guide fields and plasma $\beta \sim 1$. The variation of important observational properties during this transition remain unknown, i.e., the existence of frozen-in ion outflows, the ion outflow speed, and the width along the normal direction of the ion exhaust. 

In this paper, we study the transition from ion-coupled to electron-only reconnection using kinetic particle-in-cell (PIC) simulations of magnetic reconnection. The initial inflow conditions for the simulation are relevant for turbulent reconnection in the magnetosheath, i.e., relatively large plasma $\beta$ and weak magnetic shear. We simulate varying simulation domain sizes and examine the effect on the ion response to the reconnection. We find that the transition between fully ion-coupled and electron-only reconnection is gradual, spanning nearly a factor of ten in domain size. This transition is characterized by a gradual increase in the ion outflow velocity, the ion out-of-plane current, and the degree to which the ions are frozen-in to the magnetic field. Electron-only reconnection exhibits much faster reconnection rates because the magnetic field motion is not limited by the Alfv\'en speed. We develop a simplistic model for a newly reconnected field line which accurately predicts the scaling of peak ion outflows with domain size. A key finding is that the ion outflow velocity is largely controlled by the exhaust width along the current sheet normal direction.

We then explore the implications of our findings. First, the relationship between exhaust width and ion response gives specific predictions for both ion outflow speeds and ion out-of-plane current that can be compared with observations. Second, we examine how the properties of turbulence impact the degree of ion coupling in the resultant reconnection.

Note that a terminology issue arises in the simultaneous analysis of laminar reconnection simulations and reconnection as an element of turbulence. The magnetic flux structures currently undergoing reconnection have been variously called ``magnetic flux bundles"~\cite{Shay98a}, ``unreconnected magnetic islands"~\cite{Matthaeus86}, and possibly other names. The flux structures consisting of already reconnection magnetic field lines have been called ``magnetic islands", ``reconnected magnetic islands", ``magnetic bubbles", as well as ``plasmoids". To avoid confusion here, we will use the term ``magnetic bubbles" to describe magnetic flux structures currently undergoing reconnection and ``magnetic islands" for flux structures composed of already reconnected magnetic field. We emphasize that the use of the term ``bubble'' does not imply that the reconnection structures are small. In our usage a ``bubble'' could have a diameter of thousands of ion inertial lengths. 

Section~\ref{simulations} describes the simulations performed in this study. In section~\ref{sim_and_results}, the simulation results and analyses are presented and the model for ion outflows is described. Section~\ref{implications-observations} discusses the implications for observational signatures of reconnection. Section~\ref{implications-turbulence} discusses how our findings impact our understanding of ion coupling to reconnection in turbulence. Finally, in Section~\ref{conclusions} we review and discuss our scientific results.


\section{Simulations}\label{simulations}

To study the physics of small-scale magnetic reconnection relevant to the turbulent magnetosheath (plasma $\beta \gtrsim 1$ and large guide field), we have performed 6 different simulations described in Table~\ref{plasmaparameters} using the multi-parallel particle-in-cell (PIC) code P3D~\cite{Zeiler02}. The simulations are 2.5 dimensional with periodic boundary conditions. Multiple system sizes, while keeping the same aspect ratio, are used to examine the transition from ion-coupled to electron-only reconnection.  Calculations are presented in normalized units: the magnetic field to $B_\mathrm{0}$,
density to $n_\mathrm{0}$, lengths to ion inertial length $d_i \equiv c/\omega_{pi}$,
times to inverse ion cyclotron frequency $\Omega_\mathrm{i}^{-1}$ defined
in terms of $\frac{m_i c}{e B_\mathrm{0}}$, velocities to the Alfv\'en speed $c_{A0},$ temperature to $m_\mathrm{i}c_\mathrm{A_\mathrm{0}}^2$, and electric fields to $E_\mathrm{0}=B_\mathrm{0}c_{A_\mathrm{0}}/c$. Using
the simulation normalized units, various key physical length scales can be
calculated from code values as: ion inertial length $d_i = \sqrt{1/n}$; electron
inertial length $d_e = \sqrt{(m_e/m_i)/n}$; ion Larmor radius $\rho_i = \sqrt{T_i}/B$;
and electron Larmor radius $\rho_e = \sqrt{T_e\,(m_e/m_i)}\,/\,B$.  The simulations have a domain size $L_\mathrm{x} \times L_\mathrm{y}$ and grid scale $\Delta$.

The simulations are initialized with two current sheets, with the magnetic field along $x$ given by $B_\mathrm{x} = B_\mathrm{up}\; \{ \; \tanh [(y-0.25\,L_\mathrm{y})/w_\mathrm{0}] - \tanh [ (y-0.75L_\mathrm{y})/w_\mathrm{0}]-1 \}$, where $w_\mathrm{0}$ is the half-width of the initial current sheets and $B_\mathrm{up}$ is the inflowing reconnecting magnetic field. $n_\mathrm{up}$ is the density outside the current sheets and the density is varied to maintain total pressure balance.  A small local magnetic perturbation is added to start the reconnection, and the initial currents are due solely to electron flows. Run~A, B, C, D, and E have 6000 particles per grid (ppg) in the regions outside the current sheets, while Run~F has 1500~ppg. The lower ppg for run F was necessary to prevent the simulations from being too computationally expensive. Temperatures are initially uniform and there is also an initial uniform large guide field $B_z = B_g$. Parameters for the simulations are shown in Table~\ref{plasmaparameters}. The inflow conditions are similar to ~\citet{Phan18}. Note that because $B_\mathrm{up} = 1$ and $n_\mathrm{up} = 1$, velocities and reconnection rates are normalized to the inflowing Alfv\'en speed $c_\mathrm{Aup}$. Lastly, the simulation sizes of Run A through F are notated interchangeably by their domain sizes: $\mathrm{2.5d_i}$, $\mathrm{5d_i}$, $\mathrm{10d_i}$, $\mathrm{20d_i}$, $\mathrm{40d_i}$, and $\mathrm{80d_i}$. The domain sizes are used where the use of length scales are deemed instructive.

\begin{table*}
\caption{
Plasma parameters of six simulations (Runs) : $m_i/m_e$ is the mass ratio of ion to electron and $B_\mathrm{up}$ and $n_\mathrm{up}$ are the inflowing reconnecting field and density outside the current sheet, respectively. $B_g$ is the uniform guide field, $\Delta$ is grid scale, $c$ is light speed, and ($L_\mathrm{x}$, $L_\mathrm{y}$) are simulation domain sizes. $\beta$ is the total beta including the guide field. $T_e$ and $T_i$ are the uniform electron and ion temperatures. $w_0$ is the initial current sheet thickness.}
\begin{tabular}{| c | c | c | c | c | c | c | c | c | c | c | c | c |}      
        Run  & $\frac{m_i}{m_e}$ & $B_\mathrm{up}$ & $n_\mathrm{up}$ & $T_\mathrm{e}$ & $T_\mathrm{i}$ & $B_{g}$ & $c$ & $L_\mathrm{x}$ & $L_\mathrm{y}$ & $\Delta$ & $\beta$ & $w_0$\\   
        \hline          
        A & 1836    & 1   &  1 & 11.51     & 115.16    & 8  & 300 & 2.56 & 2.56 & 0.005 & 3.89 & 0.06  \\
        B & 1836    & 1   &  1 & 11.51     & 115.16    & 8  & 300 & 5.12 & 5.12  & 0.005 & 3.89 & 0.04  \\
        C & 1836/16 & 1   &  1 & 11.51     & 115.16    & 8  & 100 & 10.24 & 10.24 & 0.02 & 3.89 & 0.065 \\
        D & 1836/16 & 1   &  1 & 11.51     & 115.16    & 8  & 100 & 20.48 & 20.48 &  0.02 & 3.89 & 0.22 \\
         E & 1836/64 & 1   &  1  & 11.51   & 115.16    & 8  & 50 & 40.96 & 40.96 &  0.035 & 3.89 & 0.4 \\
         F & 1836/64 & 1   &  1  & 11.51   & 115.16    & 8  & 50 & 81.92 & 81.92 &  0.04 & 3.89 & 0.6 \\
\end{tabular}
\label{plasmaparameters}
\end{table*}

\section{Simulation Results and Discussion}\label{sim_and_results}

\begin{figure*}  
\includegraphics[width=0.83\columnwidth]{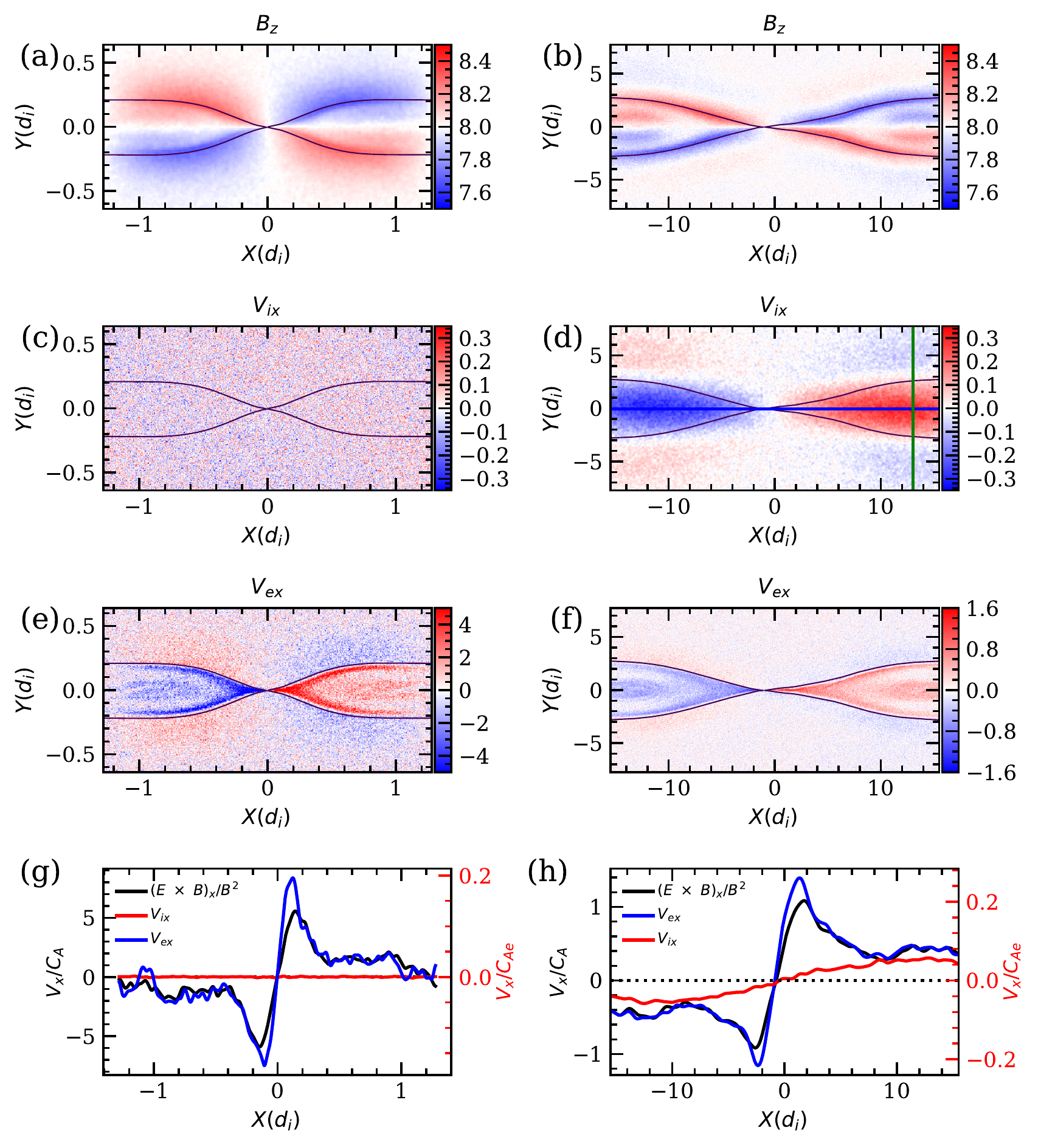}
\caption{
Overview of the reconnection simulations: time slice taken at $t=0.265\Omega_i^{-1}$ for $2.5d_i$ (left column) and at $t=27\Omega_i^{-1}$ for $40d_i$ (right column). (a) The quadrupolar structure of the out of plane magnetic field. (b) The out of plane magnetic field resembles that of a typical magnetic reconnection. (c) No ion exhaust velocity $V_{ix}$ is observed. (d) Significant ion outflows $V_{ix}$ are present. The intersection between the blue horizontal line and the green vertical line is the location of the maximum value of $V_{ix}$. (e) Electron outflow  $V_{ex}$. The electron diffusion region is characterized by very fast collimated electron outflows near the x-line.  (f) Peak electron jet close to the electron diffusion region is larger than the ion jets in (d). (g) A cut along $x$ at $y = 0$ is taken along the reconnection mid-plane. The electron outflow and $\mathrm{E \times B}$ drift are very similar and $V_{ix}$ shows no ion response. (h)  A cut along $x$ at $y = 0$ is taken along the reconnection mid-plane. Outside the diffusion region, $V_{ex}$ and $\mathrm{E \times B}$ drift decrease slowly to match $V_{ix}$ at $\approx 10d_i$. The ions have fully coupled in this simulation.} 
\label{fig:overview_plot}
\end{figure*}

An overview of the reconnection simulations are shown in Figure~\ref{fig:overview_plot}; the left column is the smallest simulation domain (Run A) and the right column is Run E. 

The larger simulation exhibits standard ion-coupled reconnection (Figure~\ref{fig:overview_plot}b,d,f), with a quadrupolar $B_z$ perturbation, an ion outflow exhaust, and an electron flow characterized by super-Alfv\'enic flow close to the x-line, and then exhaust flows similar to the ions farther downstream. In contrast, the smallest simulation (Figure~\ref{fig:overview_plot}a,c,e) exhibits a quadrupolar $B_z$ perturbation that extends beyond the current sheet, negligible ion outflow, and electron outflows peaked near the separatrices. Due to the lack of ion response, we follow~\citet{Phan18} and call this ``electron-only'' reconnection. The $B_z$ perturbation which fills the inflowing region in this reconnection is generated in part by the electron inflow which by necessity is a current.

Shown in Figure~\ref{fig:overview_plot}g,h are cuts of the outflow velocities along the midplane ($y = 0$) compared to the $\mathbf{(E \times B)}_x/B^2$ $(\mathrm{E\times B} \ \mathrm{drift})$, which reveal the electron and ion coupling explicitly. In the electron-only case there are no ion flows and the electron outflow follows the $\mathrm{E\times B}$ drift velocity closely. Note that for both simulations, this strong guide field reconnection has a significant $E_\parallel$ close to the x-line, so the electrons are not frozen-in there even though $V_{ex} \approx \mathbf{(E \times B)}_x/B^2$. For the ion-coupled reconnection in Figure~\ref{fig:overview_plot}h, the electron flows reach velocities much greater than the ions close to the x-line, and then slow down to roughly match the ion flows approximately $\mathrm{10\,d_i}$ downstream of the x-line. The ions become frozen-in at this location with $V_{ix} \approx \mathbf{(E \times B)}_x/B^2$. 

Due to the computational cost of larger simulations, smaller $m_i/m_e$ are used. The result shows that the electron flows for electron-only reconnection are much greater than those for ion-coupled when normalized to the ion Alfv\'en speed (Figure~\ref{fig:overview_plot}g,h). However, when the electron flows are normalized to the electron Alfv\'en speed $c_{Ae}$, the peak electron flows are nearly identical; to highlight this fact, the right axes show values for $V_{x}/c_{Ae}$. As the ions couple more fully to the reconnection with increasing system size, the reconnection rate is lower because magnetic flux cannot flow away from the x-line as quickly as in the electron-only case. 
\begin{figure}  

\includegraphics[width=0.5\columnwidth]{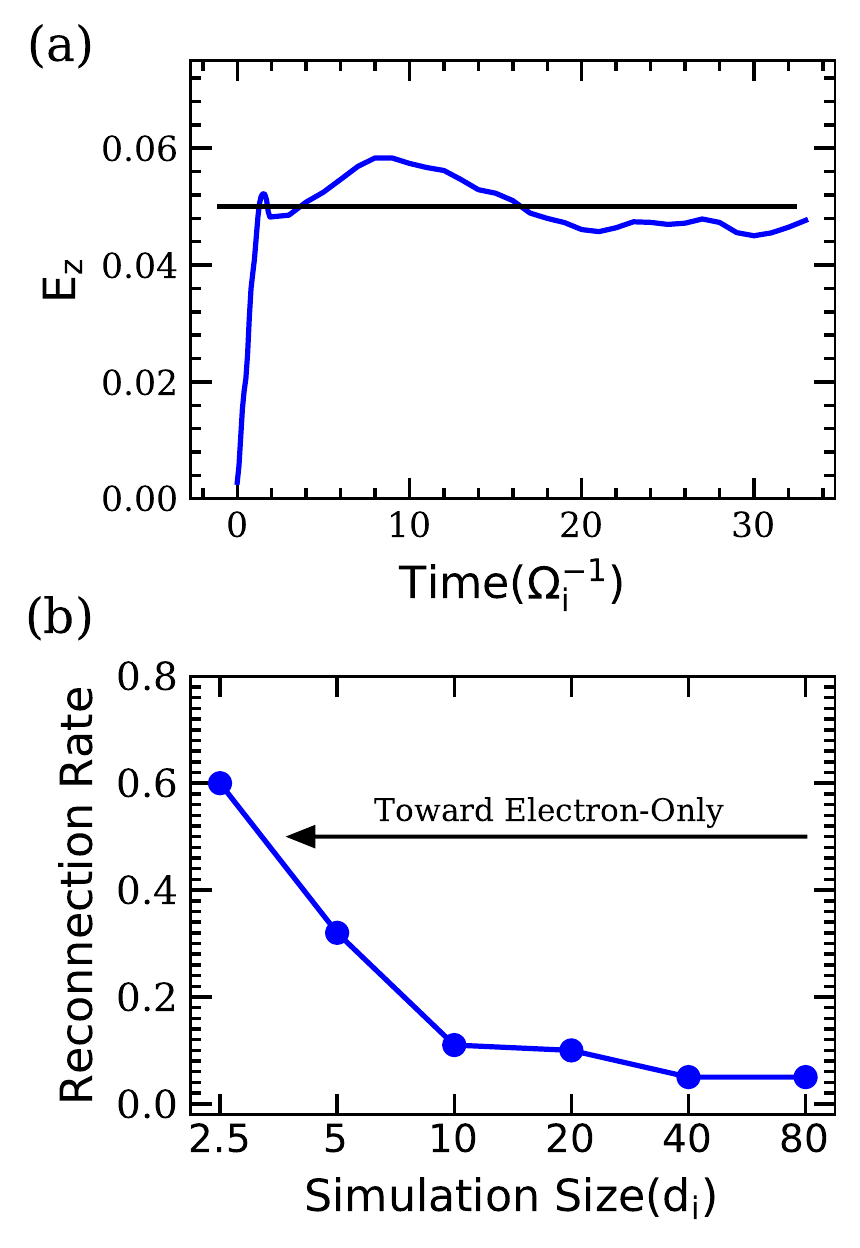}
\caption{
(a) Run~E: Reconnection rate vs. time. The straight black line is the steady-state reconnection rate of 0.05 and the peak value is 0.06. (b) Reconnection rate vs. system size for all the simulations: Electron-only reconnection has a reconnection rate significantly larger than the ion coupled reconnection rates. Notably, the reconnection rate converges to 0.05 as MHD scales are realized. 
\label{fig:rrates}}
\end{figure}

The reconnection rate $E_z$ is calculated by taking the time derivative of the magnetic flux between the x-point and the o-point. The reconnection rate $E_z$ versus time for the $40d_i$ simulation (Run~E) is shown in Figure~\ref{fig:rrates}a. The reconnection rate rises and asymptotes to a value drawn by the horizontal black line in Figure~\ref{fig:rrates}a. The effect of system size on the quasi-steady value is shown in Figure~\ref{fig:rrates}b. As the simulation domain is increased from the smallest size, initially the quasi-steady reconnection rate decreases, an effect which has been found in previous hybrid~\cite{Mandt94} and kinetic PIC simulation studies~\cite{Shay98b}. Both of these studies determined that for smaller system sizes, whistler physics, as opposed to MHD, was controlling the reconnection rate. For larger systems, the reconnection rate stabilizes to a value consistent with previous reconnection scaling studies~\cite{Shay99,Birn01}. 

An important aspect of the transition between ion-coupled and electron-only reconnection that has not been previously addressed is the onset of ion flows. Clearly, as reconnection proceeds, and if the reconnection geometry extends to scales much greater than the ion Larmor radius, the ions will fully couple as in Run~E and F. However, is this transition sudden or does it gradually occur? What controls the onset? To address these questions, we study the ion flow properties as the system size is increased.

\begin{figure*} 
\includegraphics[width=0.85\columnwidth]{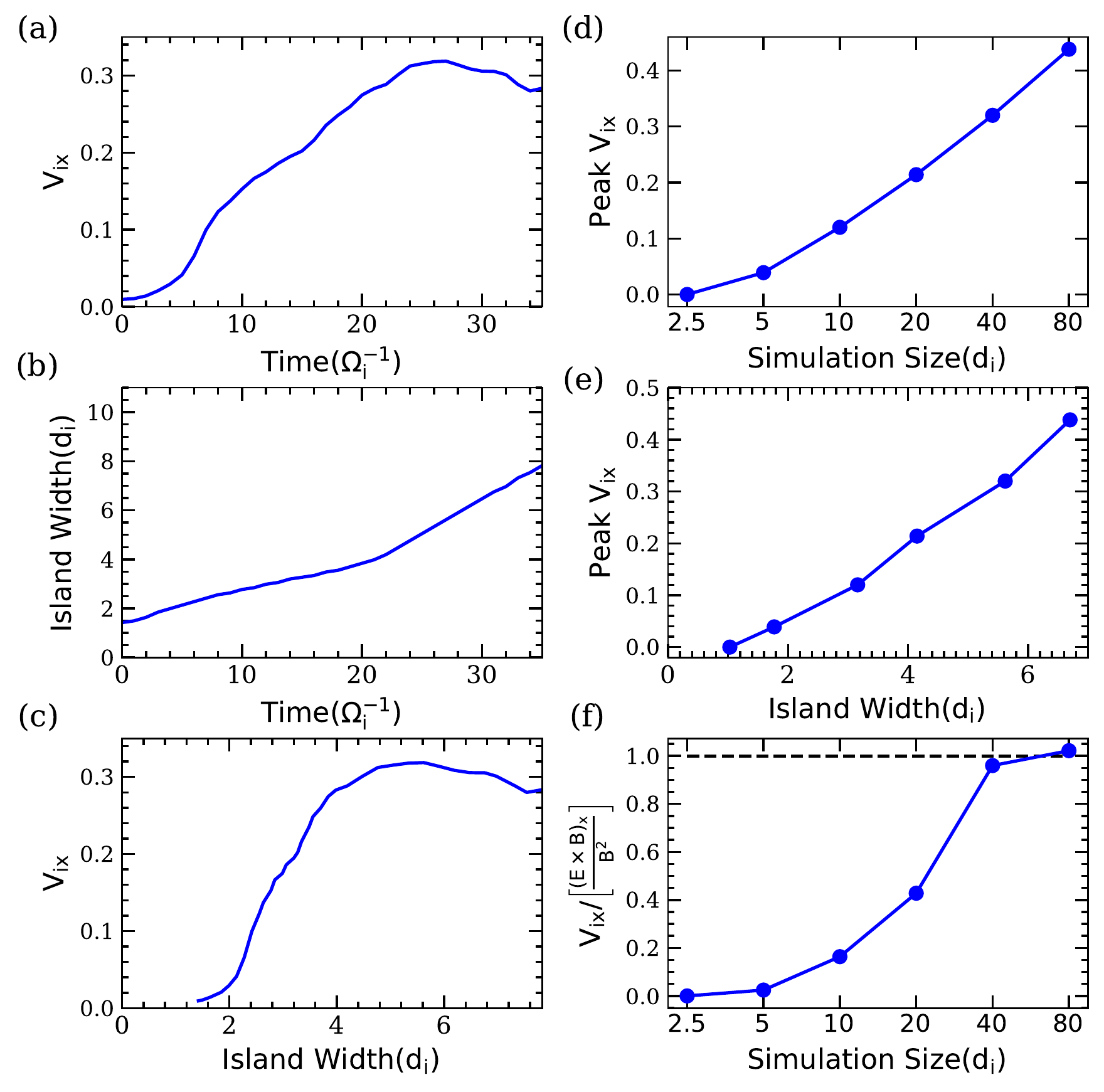}
\caption{
 (a) Run E: Outflow speed vs. time. The outflow velocity is measured at the intersection of the horizontal blue and the vertical green line shown in Figure~\ref{fig:overview_plot}d for each time slices. The peak outflow velocity of about $\sim 0.32$ occurs at $t=27$. (b) Run E: Total island width vs. time. The separatrices associated with the primary x-line form the boundary of the magnetic island. The total island width is the normal distance between the separatrices at the O-line. (c) Run E: Outflow velocity $V_{ix}$ versus island width. The peak value 0.32 is attained when the size of the island width is about $\sim 5.6\mathrm{d_i}$. (d) All runs: Peak $V_{ix}$ vs. system size of simulations.  (e) All runs: Peak $V_{ix}$ vs. island width of simulations. The island width is measured at the time when the outflowing velocity $V_{ix}$ has peaked.  (f) All runs: Ratio of the peak ion outflow $V_{ix}$ and $\mathrm{E\times B}$~drift at the midplane. As the simulation size gets bigger, the ion outflow gradually reaches the $\mathrm{E\times B}$~drift speed, indicating full ion coupling. 
\label{fig:R_rate_peak_vix_time}}
\end{figure*}

To characterize these ion flows, we begin with a cut along $x$ of $V_{ix}$ at the midplane of the exhaust in Run E, an example of which is shown in Figure~\ref{fig:overview_plot}h. Note that we average $0.1(c/\omega_{pi})$ above and below the midplane and a 1d-Gaussian filter is applied with a width corresponding to $0.07(c/\omega_{pi})$. Gaussian filtering proves to be an effective tool to reduce noise~\cite{Haggerty17}. The peak outflow speed along this cut at $t \approx 27 \Omega_{i}^{-1}$ is $|V_{ix}| \approx 0.32$ to the right of the x-point. The location of this peak outflow speed is also shown by the intersection of red vertical and blue horizontal line in Figure~\ref{fig:overview_plot}d. The peak outflow speed at each time is determined similarly and time evolution of this peak outflow speed is shown in Figure~\ref{fig:R_rate_peak_vix_time}a. The outflow speed rises in time and reaches an apex of 0.32 and descends. We choose this apex value as the characteristic outflow speed for a given system size (simulation) and plot the results for each simulation in Figure~\ref{fig:R_rate_peak_vix_time}d. It is clear from this figure that the characteristic ion outflow speed smoothly increases with system size. 

The ion outflow velocity grows with system size because the ions can only fully couple to the reconnection process when the exhaust region is significantly larger than the ion Larmor radius, which is $\mathrm{1.34~d_i}$ for these simulations. The maximum width of the exhaust along the normal direction can be estimated as the total magnetic island width, which is shown for Run~E in Figure~\ref{fig:R_rate_peak_vix_time}b; this width grows steadily in time as the reconnection proceeds. The ion outflow does not reach its characteristic speed until the total island width is a few ion Larmor radii in width ($t \approx 27$), as shown in Figure~\ref{fig:R_rate_peak_vix_time}c. Intuitively then, the smaller simulation domains simply do not allow the magnetic island to become large enough to allow the ions to fully couple to the magnetic fields in reconnection, resulting in lower ion outflow velocities. This fact is highlighted in Figure~\ref{fig:R_rate_peak_vix_time}e, which shows the characteristic ion outflow velocity versus the total island width for each simulation when this characteristic ion outflow speed is reached. It is clear from this figure that the transition to ion-coupled reconnection is gradual and not sudden, with the characteristic ion outflow speed smoothly increasing with system size. Further, in Figure~\ref{fig:R_rate_peak_vix_time}f, we show the ratio of $V_{ix}$ and $\mathrm{E\times B}$ drift for each simulation at its peak. As the system size increases, the characteristic ion outflow velocity catches up to the $\mathrm{E\times B}$ drift velocity: another clear indication that the ions have fully coupled for the larger system sizes. 


To understand more quantitatively the physics behind this transition from ion-coupled to electron-only reconnection, we study the physics controlling the contraction of a strongly-curved newly-reconnected field line by approximating this field line as a linear wave; note that in the limit of $k\,\mathrm{d_i} \ll 1$ the field line is approximated as an MHD Alfv\'en wave.  Although reconnection is a nonlinear phenomenon, this type of analysis has previously been used successfully to predict the electron outflow speed at sub-MHD length scales \cite{Shay98b,Shay01,Cassak10} and to study the propagation and damping of the Hall magnetic fields generated during reconnection \cite{Shay11,Pyakurel2018}. It has also been used to motivate why the global reconnection rate is ``fast" or  independent of the dissipation mechanism and system size \cite{Shay99,Rogers01}, but this conclusion has been the source of significant and ongoing controversy (e.g., ~Refs.~\onlinecite{Bessho05,Daughton07,Chacon08,TenBarge14,Liu14}). In this study we exclusively focus on using this type of model to give predictive insight into the ion reconnection exhaust velocity and we find that linear theory successfully predicts the scaling of this velocity. 

The predicted ion outflow velocity is the bulk ion flow speed generated by the wave, which in the MHD limit becomes the Alfv\'en speed based on the inflowing plasma conditions. The wavevector $\mathbf{k}$ is taken to be along $y$ with the background field $B_0 = \sqrt{B_y^2 + B_z^2}$, where $B_z$ is the guide field $B_g$ in Table~\ref{plasmaparameters}; $B_y$ is chosen to be the value at the location of peak ion outflow, which is 0.18 for Run~E. The angle of propagation relative to the background field is $\theta = \tan^{-1} (B_z/B_y)$ and the wave is obliquely propagating. The perturbation field is $B_x$, which is $B_\mathrm{up}$ in Table~\ref{plasmaparameters}; $B_\mathrm{up} = 1$ for all of the simulations.

\begin{figure} 
\includegraphics[width=0.6\columnwidth]{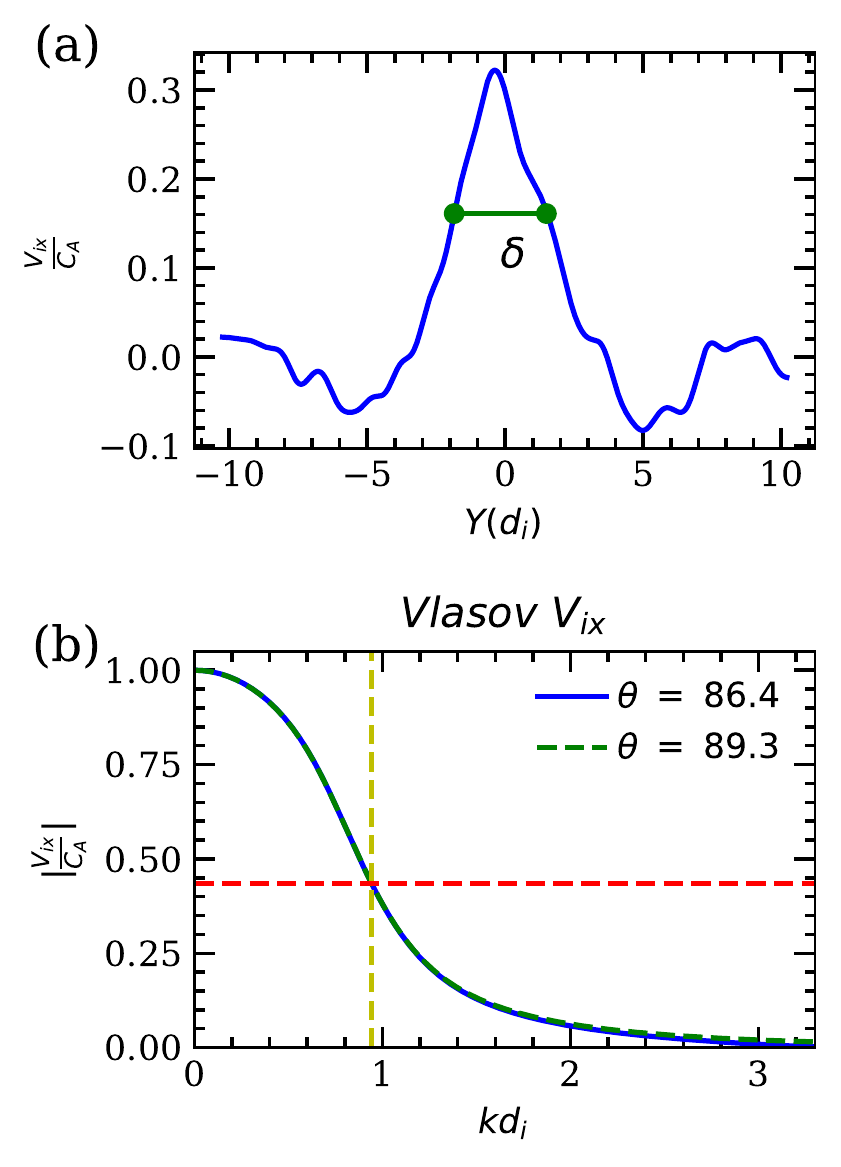}
\caption{Run E: Determining the theoretical prediction of ion outflow velocity for run
E at $t = 27\Omega_i^{-1}$ (a) Slice along $y$ of $V_{ix}$ at the location of peak $V_{ix}$
( $x=13.04$ in Figure~\ref{fig:overview_plot}d and h). The width $\delta$ gives
$k = (2\pi/2\delta)$. (b) $V_{ix}$ versus $kd_i$ from a numerical Vlasov
dispersion solver~\cite{Klein15}; 
two different angles of propagation $\theta = \tan^{-1} (B_z/B_y)$ are shown,
corresponding to $B_y$ = 0.5 and 0.1.
The dashed yellow vertical line shows $kd_i$ value determined from (a), giving
the theoretical prediction for $V_{ix}$ (dashed red line). 
\label{fig:wave-properties}}
\end{figure}

To determine the magnitude of $\mathbf{k},$ we examine the width $\delta$ of the ion $V_x$ along $y$ as shown in Figure~\ref{fig:wave-properties}a.
The cut is taken at the location where the ion $|V_{ix}|$ is peaked to the right of the x-line in Figure~\ref{fig:overview_plot}d, which is denoted by the vertical green line. The width $\delta \approx 3.3d_i$, which is the full width at half maximum, is converted to a wave number using $k \approx \frac{2\pi}{2\delta} \approx 0.94d_i^{-1}$.

Numerical solutions for the linear dispersion relation were calculated using the PLUME numerical solver~\cite{Klein15}. For a set of equilibrium background parameters, in this work $\beta_i$, $T_i/T_e$, and $v_{th_{i}}/c$, PLUME determines the normal mode frequency $\frac{\omega}{\Omega_i}$ solutions of the hot plasma dispersion relation as a function of wavevector $\mathbf{k}d_i$, using a full Bessel Function representation of the ions and electrons as well as the associated eigenfunction fluctuations, e.g, the ion velocity flow shown in Figure~\ref{fig:wave-properties}b. As $k$ increases, the ion coupling to the wave decreases leading to a slower $V_{ix}$. The ion velocity has little dependence on $\theta = \tan^{-1} (B_z/B_y)$ for these oblique angles. The two angles shown correspond to $B_y = 0.5$ and $0.1$ but the two curves almost completely overlap. The dashed yellow vertical line denotes the value determined from Figure~\ref{fig:wave-properties}a, $k \approx 0.94d_i^{-1}$, giving
the theoretical prediction for $V_{ix} \approx 0.43$ (dashed
red line) shown in Figure~\ref{fig:wave-properties}b.

\begin{figure}
\centering
\includegraphics[width=0.7\columnwidth]{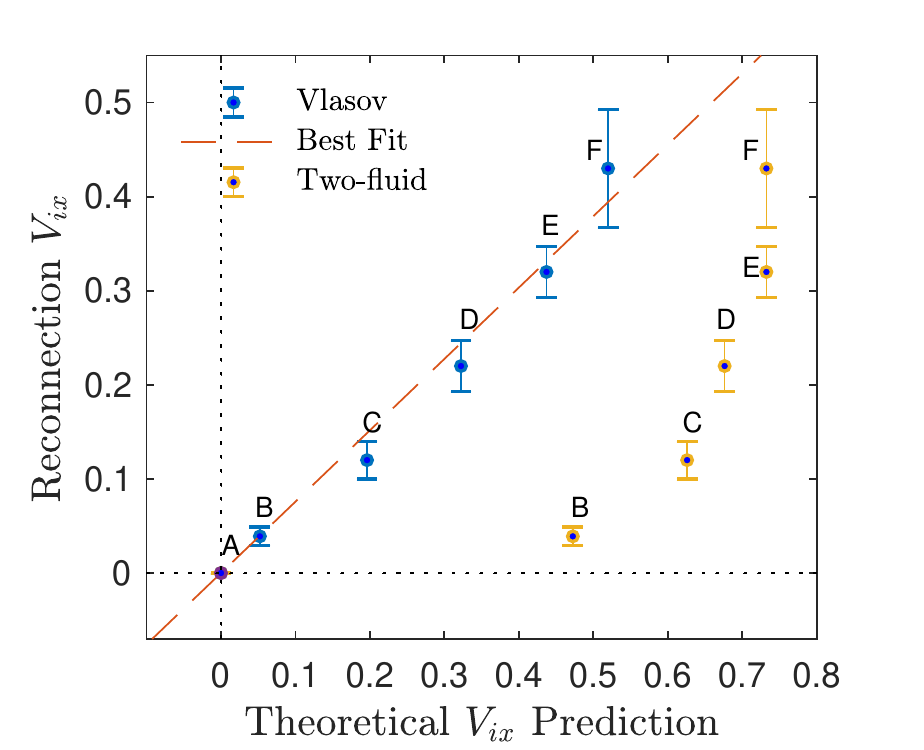}
\caption{
Comparison of the peak reconnection ion outflow velocity $V_{ix}$ with theoretical
predictions. Theoretical predictions using both two-fluid and Vlasov dispersion
relations are shown. The dashed red line with slope = 0.75 is the best fit
line for the Vlasov prediction. The reconnection $V_{ix}$ are averaged values of $V_{ix}$
once they peak in each simulation. For example, Run~E is measured from  Figure~\ref{fig:R_rate_peak_vix_time}a when $t \approx 27\Omega_{i}^{-1}$. The uncertainty in $V_{ix}$ is estimated from the standard deviation of the fluctuations in time. 
\label{fig:outflow-prediction}}
\end{figure}

For all of the simulations in this study, a comparison of the measured versus theoretical predictions for the peak ion outflow are shown in Figure~\ref{fig:outflow-prediction}.
The Vlasov prediction organizes the data in a straight line with a slope of approximately 0.75, shown as the dashed red line. The slope is calculated using simple linear regression.  For contrast, we also include a prediction from the isothermal two-fluid theory from which the calculation of the eigenvectors is straightforward (see ~Refs. ~\onlinecite{Formisano69} and ~\onlinecite{Rogers01}). Clearly due to the relatively high ion $\beta$, finite ion Larmor radius effects are playing an important role in the ion response to the reconnection. Note that the error bars for run F are significantly larger than the other simulations because of the lower particles-per-grid used.

\section{Implications for Reconnection Observations} \label{implications-observations}

\begin{figure*}
\centering
\includegraphics[width=0.85\columnwidth]{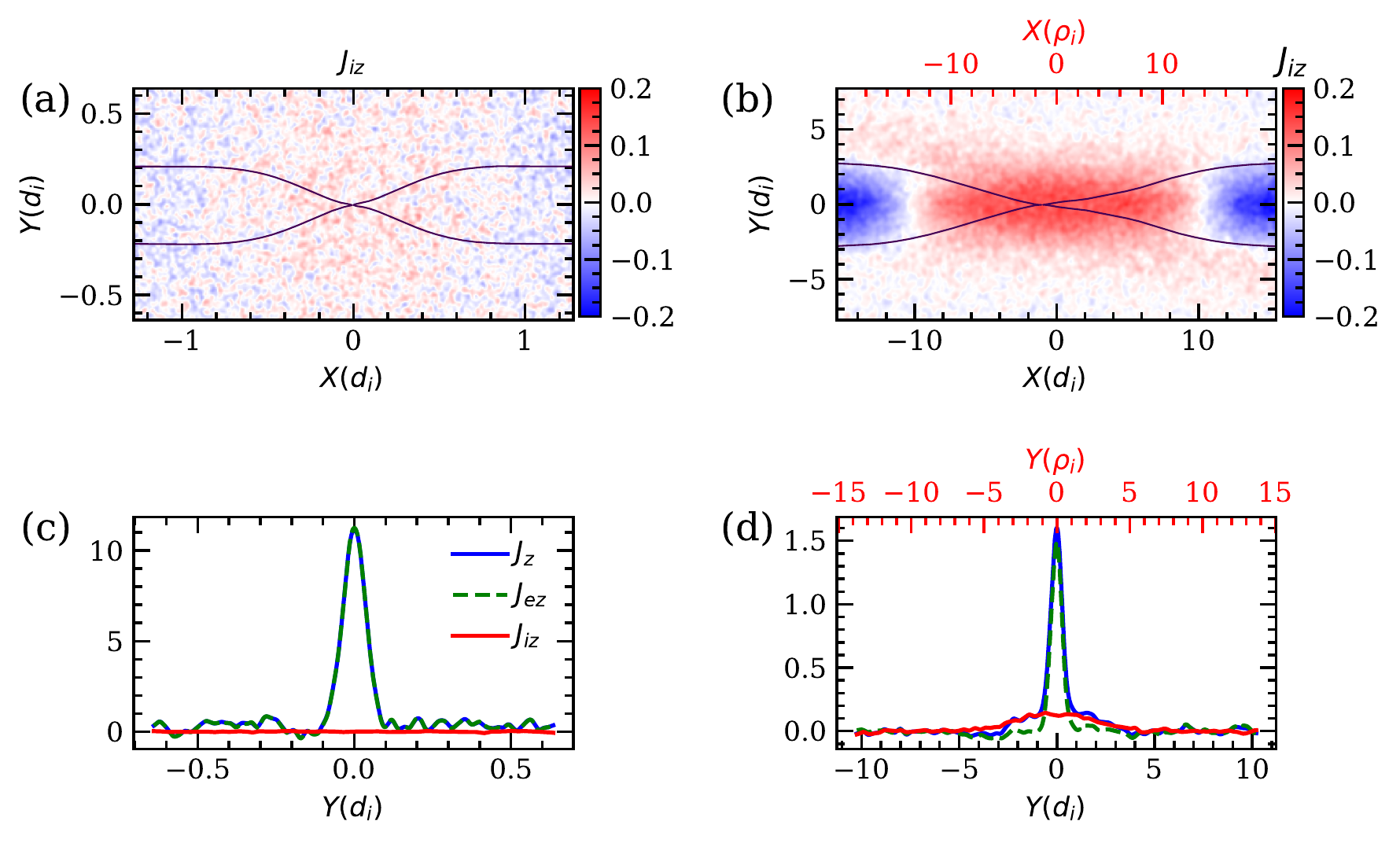}
\caption{
Time slice taken at $t=0.265\Omega_i^{-1}$ for $2.5d_i$ (left column) and at $t=27\Omega_i^{-1}$ for $40d_i$ (right column). (a) No out of plane ion current $J_{iz}$ is observed. (b) Significant out of plane ion currents $J_{iz}$ are present. (c) A cut along $y$ is taken at the location of peak $V_{ex}$ ($x=0.11$ in Figure~\ref{fig:overview_plot}g). (d) A cut along $y$ is taken at the peak location of $V_{ex}$ ($x=1.31$ in Figure~\ref{fig:overview_plot}h). \label{fig:Jiz_cuts}}
\end{figure*}

Recent MMS observations of the turbulent magnetosheath~\cite{Phan18} found smoking gun evidence for magnetic reconnection in the form of diverging super-Alfv\'enic electron plasma jets. The event was novel because it showed electron-only reconnection without ion coupling. First, the reconnection current sheet showed no evidence of the two-scale structure typical of ion-coupled reconnection~(Ref.~\onlinecite{Shay98a} Figure 3), i.e., a weaker ion-scale current sheet and an intense electron scale current sheet. Second, the ions showed no change in their velocity due to the reconnected magnetic field lines. Additionally, no ion flows were observed in any of the current sheets that were observed. The simulations performed in this study have plasma inflow conditions often found in the downstream of a quasi-parallel bow shock in the magnetosheath (relatively high $\beta$, significant guide field), and can therefore provide some context for interpreting observations.

First, the transition from a two-scale ion-coupled sheet to an electron-only reconnection current sheet is evident in the simulations. Figure~\ref{fig:Jiz_cuts}a and b show the Gaussian filtered ion out-of-plane current $J_{iz}$ for runs A and E, with both having the same color scale. Note that $|J_{iz}| \approx |V_{iz}|$ in this study because the density is nearly constant with a value of $1.0$ because the flows are low Mach number. While run A shows no ion response, in run E the ions have a rectangular current sheet typically seen in ion-coupled reconnection~\cite{Shay98a}. This ion current sheet extends almost $\mathrm{10\,d_i}$ downstream from the x-line. 

A spacecraft crossing the electron dissipation region in these two cases will see very different structures. In Figure~\ref{fig:Jiz_cuts}c and d, we plot electron, ion, and total currents in a cut along $y$ through the location of peak electron outflow, i.e., near the outflow edge of the electron diffusion region. This smoothed cut is located at $x=0.1$ and $x=1.31$ for runs A and E, respectively. In the electron-only case, the only current comes from an electron current sheet with a total width of roughly $\mathrm{8\,d_e \approx 0.2\,d_i}$. In contrast, the ion-coupled case (run E) exhibits an ion current sheet of width of approximately $ 8\,\mathrm{\rho_i=8}\, \sqrt[]{\frac{\beta_i}{2}}\mathrm{d_i}$; because $\beta_i \gtrapprox 1,$ the ion current sheet width is set by the ion Larmor radius instead of the ion inertial length. As with $V_{ex}$ in Figures~\ref{fig:overview_plot}g and h, the electron currents are smaller in run E because they roughly scale with $\sqrt{m_i/m_e}.$ 

\begin{figure*}
\centering
\includegraphics[width=0.85\columnwidth]{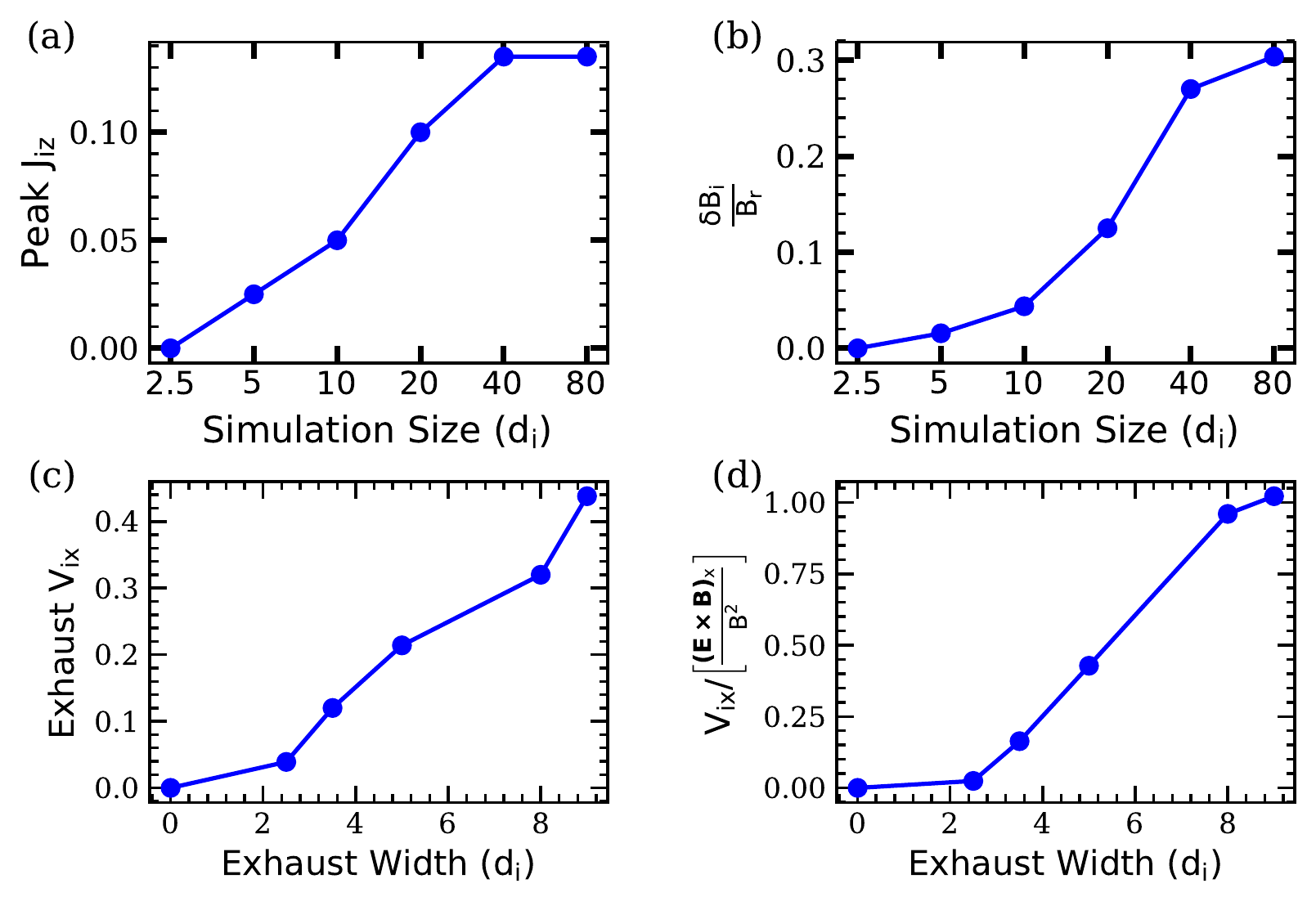}
\caption{All runs: each point represents a simulation. (a) Peak $J_{iz}$ gradually increases and plateaus at simulation size of $40 d_i$. (b) Reduction in the inflowing reconnecting magnetic field for given simulations. (c) The peak ion exhaust velocities are plotted against the exhaust widths for given simulations. A gradual increase is seen. (d) The ion exhaust speeds normalized to $\mathrm{E\times B}$ drift velocity for each simulation are plotted against the exhaust width. \label{fig:Jiz_size}}
\end{figure*}

Similar to $V_{ix},$ the transition between electron-only and ion coupled $J_{iz}$ signature is gradual as the system size increases. In Figure~\ref{fig:Jiz_size} is shown the peak value of $J_{iz}$ in each simulation plotted versus system size, showing a gradual increase in $J_{iz}$ with system size until a plateau is reached for the largest two simulations. A cut along the mid-plane is taken and an average peak value is inferred from this cut to determine $J_{iz}$. Generally, the peak value of $J_{iz}/n_i$ in simulation normalized units is roughly half the peak value of $V_{ix}$ when compared with Figure~\ref{fig:R_rate_peak_vix_time}d. 

The larger scale ion current sheet causes a gradual reduction over ion scales of the reconnection magnetic field in the inflow region. The expected change in the magnetic field due to the ion current is calculated by integrating the ion current from deep in the inflow region to the center of the current sheet: 
\begin{equation}
\delta B_i  = \int_{-\mathrm{inflow}}^0 dy\,J_{iz} \approx \frac{1}{4}\,\Delta_i\,J_{iz\, \mathrm{peak}},
\end{equation}
where $\Delta_i$ is the half width of the ion current sheet. For run E with $\Delta_i \approx 8\,\rho_i$, this approximation gives $\delta B_i/B_r \approx (1/4)\,(9.4)\,(0.13) \approx 0.3$. We calculate $\delta B_i/B_r$ for each simulation by directly integrating $J_{iz}$ and the result is shown in Figure~\ref{fig:Jiz_size}b. The magnetic perturbation gradually increases with system size but roughly asymptotes for the two largest systems at a value of around 0.3. 

If fully developed reconnection is strongly coupled to the ions, a significant perturbation to the reconnection magnetic field would be expected at scales of several ion inertial lengths or ion Larmor radii. Note that these scales are of order 100 times larger than the thickness of the electron diffusion region for a realistic mass ratio, so a spacecraft like MMS is very unlikely to see the ion bulk flow if crossing through the electron diffusion region. 

Another important insight from this study is that the width of the ion exhaust is linked to the peak outflow speed. The width of the ion exhaust is measured at the location of the peak ion velocity described in section \ref{sim_and_results}. In Figure~\ref{fig:Jiz_size}c we plot the peak outflow velocity for each simulation compared to the exhaust width. The ion exhaust width is measured at the location of peak ion outflow velocity. The peak velocity continues to increase up to exhaust widths of order $\mathrm{10\,d_i}$. In Figure~\ref{fig:Jiz_size}d, we plot the peak velocity normalized to the local $\mathrm{E \times B}$ drift speed. This normalized velocity increases with exhaust width, ultimately plateauing when the ions become fully coupled for exhaust widths of around $\mathrm{8\,d_i}$. If a satellite crossing the ion reconnection exhaust measures fully frozen-in ion outflow, it is expected that the exhaust width should be at least many ion inertial lengths.

\section{Implications for Turbulence} \label{implications-turbulence}

Although the simulations and analyses thus far have focused on laminar reconnection geometries, the basic relationship between exhaust width and ion coupling can be applied to turbulent systems in which reconnection can occur between adjacent interacting magnetic bubbles (flux tubes)~\cite{Servidio10}; we use the term ``magnetic bubbles'' to avoid confusion as discussed in the Introduction of this manuscript. Such an application can provide a causal linkage between turbulent length scales and the expected ion participation in subsequent reconnection. In Figure~\ref{fig:turb-schematic}a, two reconnecting magnetic bubbles (flux tubes in a 2D geometry) in a turbulent system are highlighted, where the approximate width of the exhaust $\Delta$ and diameter of the magnetic bubble $\mathrm{D}$ are shown. If the exhaust region is to have a width of at least several ion inertial lengths, it is necessary for the magnetic bubble size to be tens of ion inertial lengths. 

The constraints on magnetic bubble size in order to allow ion involvement in reconnection can be estimated using geometric arguments. A diagram of the relevant configuration is shown in Figure~\ref{fig:turb-schematic}b. Two magnetic bubbles, each of circular cross section and radius $\mathrm{r}$ interact, along the lines of what is seen in Figure ~\ref{fig:turb-schematic}a, but more simplified. Upon interaction, the boundary between the bubbles is flattened, each bubble distorted by a distance $\mathrm{\xi}$, so that a region of width $\mathrm{ \Delta =  2 \xi}$ emerges, in which the field strength drops to zero. The out-of-plane electric current density resides in this area. The flattened region defines the length $L$ of the associated reconnection zone.  On geometrical grounds we argue that the region $L$ cannot reasonably be expected to be larger than $\mathrm{r}$, as this would produce an extreme distortion and large stresses within the reconnecting flux tubes. Setting $L=\mathrm{r}$, we find by construction that $(\mathrm{r} -\xi)^2 + (\mathrm{r}/2)^2 = \mathrm{r}^2$. Throwing out a nonphysical solution with $\xi > r$, we find $\xi = (1-\sqrt{3}/2)\, \mathrm{r}$, giving a maximum value of $\Delta$ (or $\xi$) for a given bubble size $r$, namely $\xi = \Delta/2 = (1 - \sqrt{3}/2)r$. 
Consequently, to exceed a minimum specified $\xi$ requires that 
$r \gtrapprox 8 \xi $. For the minimum width $\xi$ needed for ion flows, we turn to the results of the previous sections, exemplified by 
the exhaust widths plotted in Figure~\ref{fig:Jiz_size}c.

For the particular upstream (inflow) conditions used in this study, the smallest discernible ion flow in Figure~\ref{fig:Jiz_size}c required an exhaust width of at least $\Delta \approx 2\,d_i$.  For this minimal ion participation, 
the reasoning of the previous paragraph implies 
a reconnecting magnetic bubble radius of at least $r \approx 8\,d_i$. 
Similarly, for fully ion-coupled reconnection the requirement 
is an inter-bubble separation
$\Delta \gtrapprox 8\,d_i$, which corresponds to a minimum bubble size of 
$r \approx 30\,d_i$. 

These estimates provide significant constraints on the properties of 
plasma turbulence if one anticipates that the reconnection in this turbulence is to have some degree of ion response. For large turbulence systems spanning many $d_i$ in length scales, the smallest magnetic eddies produced in the cascade may be generated at sub-$d_i$ scales (see e.g. ~Ref.~\onlinecite{Karimabadi13}). The above considerations may limit ion participation in reconnection occurring between these very small bubbles (or magnetic eddies). At the other extreme, the largest magnetic bubbles in a system are expected to be roughly the size of the turbulence correlation length, and thus the largest scale reconnection events would also occur between bubbles  of this size. Taking the threshold for minimal ion response to be $\Delta \approx d_i$ requires a magnetic bubble diameter or correlation length of at least ten ion inertial lengths. For fully coupled ions in reconnection occurring in the largest eddies, the correlation scale should be at least several 10s of ion inertial lengths. We purposely leave these constraints somewhat vague because the transition between ion-coupled and electron-only reconnection would be expected to have some dependence on inflow parameters. These estimates are consistent with recent studies of electron-only reconnection in turbulence~\cite{Califano18}, where the typical magnetic island size was less than $\mathrm{10\,d_i}$. Such reconnection would be expected to have little coupling to the ions.

\section{Conclusions} \label{conclusions}

In order to study the physics controlling the transition from fully ion-coupled reconnection to electron-only, we have performed kinetic PIC simulations of magnetic reconnection with inflow conditions appropriate  for the magnetosheath and the solar wind, i.e., plasma beta greater than $1$ and low magnetic shear. Simulations with varying domain sizes were performed to determine their effect on the reconnection rate and the ion response to reconnection, i.e., the peak ion outflow velocity, the frozen-in nature of the outflowing ions, and the generation of an ion current along the reconnection electric field (out-of-plane direction). 

For the smaller simulation domains up to about 5 ion inertial lengths, there is little or no ion response to magnetic reconnection. The magnetic field convection speed is not limited by the Alfv\'en speed which is consistent with previous studies~(e.g.~Ref.~\onlinecite{Mandt94}) and the quasi-steady reconnection rate is much faster than typical MHD-scale magnetic reconnection. As the domain size is gradually increased, the coupling of the ion flows to the reconnected magnetic field gradually increases, becoming fully coupled for a domain size of around 40 ion inertial lengths. For this domain size and larger, the quasi-steady reconnection rate asymptotes to a rate comparable to previous MHD-scale studies~(e.g.~Ref.~\onlinecite{Shay99}). The transition between electron-only and fully ion-coupled reconnection is smooth, with the ion outflows gradually becoming more frozen-in to the magnetic field as the domain size increases. The ion reconnection out-of-plane current (along the reconnection electric field) exhibits a similar gradual increase with domain size, reaching peak values of roughly one-half of the peak exhaust velocity. 

As the domain size increases, the physics controlling the ion exhaust velocity changes from kinetic Alfv\'en physics to MHD physics. We study this physics by approximating a newly reconnected and contracting magnetic field line as a portion of a linear wave~(e.g.~Ref.~\onlinecite{Drake08}; see Section~\ref{sim_and_results} for a complete discussion). The wave number of the wave is roughly inversely proportional to the reconnection exhaust width. For smaller systems with higher wave numbers, the magnetic field line acts as a kinetic Alfv\'en wave as it contracts, with little or no ion response. With larger system sizes and smaller wave numbers, the wave gradually acts as an MHD Alfv\'en wave with Alfv\'enic frozen-in ion outflows. Because of the large ion Larmor radius in the simulations, it is necessary to use a full Vlasov dispersion solver to determine the wave properties. We find good agreement between the ion outflow velocities predicted by the model and those observed in the reconnection simulations. 

We also examine how some observational signatures of reconnection vary with the degree of ion-coupling. First, an important observational clue to the degree of ion coupling has been the existence of an ion current along the out-of-plane direction surrounding the electron current sheet. As a spacecraft approaches the center of the reconnection current sheet, therefore, the reconnection magnetic field would reduce in magnitude over two different length scales. The lack of an ion scale reduction in the field (termed $\delta B_i$) provided important evidence that the ~\citet{Phan18} event was electron-only reconnection. We find that the transition between electron-only and ion-coupled reconnection is characterized by a gradual increase in the ion out-of-plane current and thus $\delta B_i$, with $\delta B_i$ ultimately reaching values of about thirty percent of the asymptotic reconnection magnetic field. 

Second, the width of the ion exhaust along the current sheet normal puts significant restrictions on both the ion flow speed and the coupling of the ions. In our simulations for a domain size of about 5 ion inertial lengths, a very small but discernible ion outflow exhaust occurred with a width of about 2 ion inertial lengths. On the other hand, to achieve frozen-in ion outflows required a minimum simulation domain of about 40 ion inertial lengths and a resultant exhaust width was about 8 ion inertial lengths. 

Finally, the link between exhaust width and ion outflow velocity has implications for our understanding of turbulence, where turbulent fluctuations lead to reconnection between magnetic bubbles. As mentioned in the introduction, to avoid confusion we call magnetic flux structures about to undergo reconnection as ``magnetic bubbles,'' and already reconnected magnetic flux structures as ``magnetic islands.'' Using geometric arguments for two reconnecting magnetic bubbles, we derive a relation between the bubble radius and the maximum reconnection exhaust width. Because the exhaust width ultimately determines the degree of ion-coupling to the reconnection, this degree can be linked to magnetic bubble size. In order to have any ion response to the reconnection, it is clear that the exhaust width must be greater than around one ion inertial length. Using our geometric relation then requires the magnetic bubble diameter to be greater than about 10 ion inertial lengths. For fully coupled ions an exhaust width $\gtrsim 5\,c/\omega_{pi}$ is required; thus, fully frozen-in ion exhausts would require a magnetic bubble size of at least several 10s of ion inertial lengths. 

We note that there is some ambiguity associated with a threshold for ``discernable'' ion flows due to reconnection. To say the least, the ability to determine if a given ion flow is associated with reconnection will depend on the global conditions driving the reconnection. A strongly turbulent system would likely have ion shear flows surrounding the reconnection site as well as significant asymmetry in inflow conditions. In our simulations we were able to discern ion outflows of around five percent of the Alfv\'en speed in the inflow region. 

Note also that the magnetic reconnection occurring in this study is well-developed reconnection, where the island width is at least $10 - 20$ electron inertial lengths. If the island width is much smaller, then the reconnection may be in a more transient onset phase. In that case the reconnection properties may be changing faster than the transit time of electrons through the diffusion region. If so, then time derivatives cannot be ignored and a Sweet Parker like analysis of the diffusion region is not applicable. This will be a topic of future research. 

\acknowledgments
This research was supported by NSF grants AGS-1219382,  AGS-1602769, AGS-1338944 and AGS-1622306 and NASA Grants 80NSSC18K015, NNX17AI25G, NNX14AC78G and NNX08A083G-MMS IDS. We would like to acknowledge high-performance computing support from Cheyenne~\cite{Cheyenne18} provided by NCAR's Computational and Information Systems Laboratory, sponsored by the National Science Foundation. This research also used resources of the National Energy Research Scientific Computing Center (NERSC), a U.S. Department of Energy Office of Science User Facility operated under Contract No. DE-AC02-05CH11231.


%
%

%


\bibliography{bib-shay.bib}

\end{document}